\documentclass{ws-procs975x65}

\begin{document}

\title{Stability of accretion disk around rotating black holes
%\footnote{\uppercase{T}his work is supported by etc, etc.}
}
\author{Banibrata Mukhopadhyay
\footnote{\uppercase{W}ork partially
supported by grant 80750 of the \uppercase{A}cademy of \uppercase{F}inland.}}
%\uppercase{N}ational \uppercase{S}cience \uppercase{F}oundation.}}
\address{Astronomy Division, P.O.Box 3000, University of Oulu,
FIN-90014, Finland}
%1060 Main Street, \\
%River Edge, NJ 07661, USA\\
%E-mail: wspc@wspc.com}

\maketitle

\abstracts{ 
I discuss the stability of accretion disks when the black hole is
considered to be rotating. I show, how the fluid properties get changed
for different choices of angular momentum of black holes. I treat the problem
in pseudo-Newtonian approach with a suitable potential from Kerr geometry.
When the angular momentum of a black hole is considered to be significant, 
the valid disk parameter 
region affects and a disk may become unstable. Also the possibility
of shock in an accretion disk around rotating black holes is checked.
When the black hole is chosen to be rotating, the sonic locations
of the accretion disk get shifted or disappear, making the disk unstable
by means of loosing entropy.
To bring the disk in a stable situation, the angular momentum of the
accreting matter has to be reduced/enhanced (for co/counter-rotating disk)
by means of some physical process.
}

%\section{}
%\subsection{Producing the Hard Copy}\label{subsec:prod}
%The hard copy may be printed using the advice given in the file
%{\em procs-readme975x65$\_$2e.pdf}, which is repeated in this section.
%Total there are six files given.\footnote{You can obtain these files
%from our WWW pages at:

The accretion phenomena around black holes is indeed a topic extensively discussed 
over the last three decades (e.g. Mukhopadhyay\cite{m03} and references therein). Here I like
to concentrate on the effect of black hole's rotation to the fundamental disk properties.
I particularly show how the fluid properties get changed with the variation of black hole's
angular momentum. Therefore the basic characteristics of a disk along with its {\it stability}
around Schwarzschild black hole may differ from that of a Kerr black hole, 
always quantitatively and in some cases qualitatively also.    

Here the word {\it stability} has a very specific meaning. We know from the second law 
of thermodynamics, entropy is the measure of disorderedness of a system. As the disorderedness
rises up, entropy increases, the system becomes more stable, and vice versa. With the change of fluid
parameters, when disks gain or loose entropy at a certain region then the stability
of that region is thought to be increased or decreased respectively. 

I will pursue the present work in the pseudo-Newtonian approach bringing a pseudo-potential
applicable for a disk around Kerr black hole. More particularly, I like to use an effective
gravitational force corresponding to a pseudo-Newtonian potential. Following an earlier work\cite{nov},
such a gravitational force was proposed\cite{m02} by me, which will be used here. An interesting point 
to note that this was a potential established for first time directly from the space-time metric.
In that respect I proposed an algorithm also, which has already been used in other works (e.g. Ghosh\cite{ghosh}).

Here the force corresponding to pseudo-potential is given as
\begin{eqnarray}
F(x)=\frac{(x^2-2a\sqrt{x}+a^2)^2}{x^3(\sqrt{x}(x-2)+a)^2}
\label{force}
\end{eqnarray}
where $a$ is specific angular momentum of a black hole and $x$ is disk radius in unit of $GM/c^2$, 
$M$ is mass of the black hole.
Throughout this paper, I also express the specific angular momentum and velocity in unit of 
$GM/c$ and $c$ respectively; others have their usual meaning. 
This $F(x)$ will be used in radial momentum balance equation to
incorporate relativistic effects approximately. 

As my main interest is to check the sole effect of black hole's rotation on the disk properties,
here I consider the inviscid fluid so that there is no dissipative force into the system and
the angular momentum of disk fluid is constant for a particular case.
The set of model equations for a disk is given as
\begin{equation}
   \frac{d}{dx}(x\rho h(x) v)=0;\hskip0.5cm
   v\frac{dv}{dx}+\frac{1}{\rho}\frac{dP}{dx}-\frac{\lambda^2}{x^3}+F(x)=0
   \label{mom}
   \end{equation} 
where, $h(x)=c_s x^{1/2}F(x)^{-1/2}$, is half-thickness of the disk and $c_s$ is sound speed.
Now following Mukhopadhyay\cite{m03}, I compute entropy, $\dot{\mu}_c\equiv\dot{M}_c$ 
($\dot{M}$ is accretion rate), and energy, $E_c$, 
at sonic points ($x_c$) from (\ref{mom}), which are useful to check stability.   
I also compute disk fluid properties solving the set of equations in (\ref{mom}), are depicted. 

\begin{figure}[ht]
%\centerline{\epsfxsize=3.in\epsfbox{f1stab.ps}}
\includegraphics[height=.25\textheight,width=0.8\textwidth]{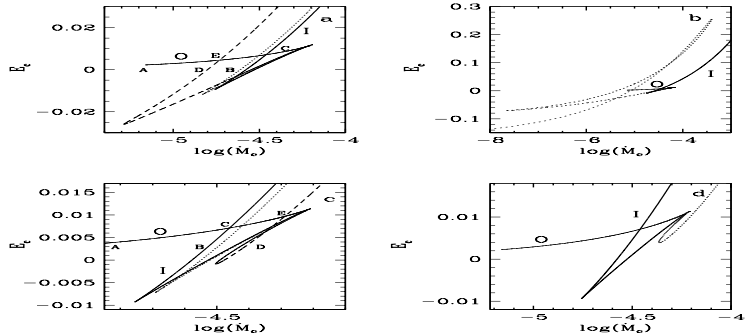}
%\plotone{f1.eps}
%\vskip-7.5cm
%\hskip3.5cm
\caption{
(a) $a=0$ (solid curve), $a=0.1$ (dotted curve) and $a=0.5$ (dashed curve), (b) $a=0$ (solid curve) and $a=0.998$
(dotted curve), (c) $a=0$ (solid curve), $a=-0.1$ (dotted curve) and $a=-0.5$ (dashed curve),
(d) $a=0$ (solid curve) and $a=-0.998$ (dotted curve). O and I are outer and inner sonic
point branches for $a=0$. $\lambda=3.3$, $\gamma=4/3$.
\label{fig1}}
\end{figure}
%\eject

First, following Chakrabarti\cite{c90} $E_c-\dot{M}_c$ analysis has been made in Fig. \ref{fig1}.
From figure it comes out that with the increase of $a$, inner sonic points form at more inside
of inner edge. But the disk looses more entropy at more inside edge, therefore for a high $a$,
disk becomes unstable compared to a situation with low or zero $a$. As the shock formation 
is directly related to the inner sonic points and its stability, for the high `$a$' shock becomes
unstable or removed (see Chakrabarti\cite{c90} for detail about shock in accretion disks). 
Also from Fig. \ref{fig1}, it is clear that with increase of $a$,
transition probability of matter from an outer sonic branch to inner one, with increase of entropy, decreases.
With increase of counter-rotation, though inner sonic points shift to stable high entropy 
region, the transition probability of matter from an outer sonic branch to inner one again decreases
and sometimes impossible (Fig. \ref{fig1}d). Therefore for both the co and counter -rotating
cases the disk and shock tend to become unstable. One should keep in mind that nothing is 
static in Universe and a rapidly rotating black hole is most natural. See Mukhopadhyay\cite{m03} for detail.

Figure 2 comes with fluid dynamical results. It directly shows that for moderately
rotating black holes, shock forms at more unstable inside region of disks with respect to that of slowly 
rotating black holes. With more increase of $a$, any shock location approaches to black hole horizon. 
Therefore for rapidly rotating black holes shock disappears completely. Mukhopadhyay\cite{m03}
discussed all these in detail.

%\begin{figure}[ht]
%\includegraphics[height=.31\textheight,width=0.7\textwidth]{f2stab.ps}
%\caption{
%(a) Mach no., (b) density and (c) temperature in unit of $10^9$
%for (i) $a=0.1$, $x_o=94$, $x_i=5.4$,
%$\lambda=3.3$ (solid curve); (ii) $a=0.5$, $x_o=43$, $x_i=4.2$, $\lambda=2.8$ (dotted curve) and
%(iii) $a=-0.1$, $x_o=70$, $x_i=6.324$, $\lambda=3.3$ (dashed curve).
%Shock invariant quantity, $C$, 
%for (d) parameter (i) when $x_s=17.18,9.2$, (e) parameter (ii) when $x_s=9.51,8.35$ and
%(f) parameter (iii) when $x_s=17.25,13.73$. Solid and dotted curves are for inner and outer sonic branch.
%$M=10M_\odot$, $\dot{M}=1$ Eddington rate, $\gamma=4/3$.
%\label{fig2}}
%\end{figure}

\begin{figure}
\includegraphics[height=.33\textheight]{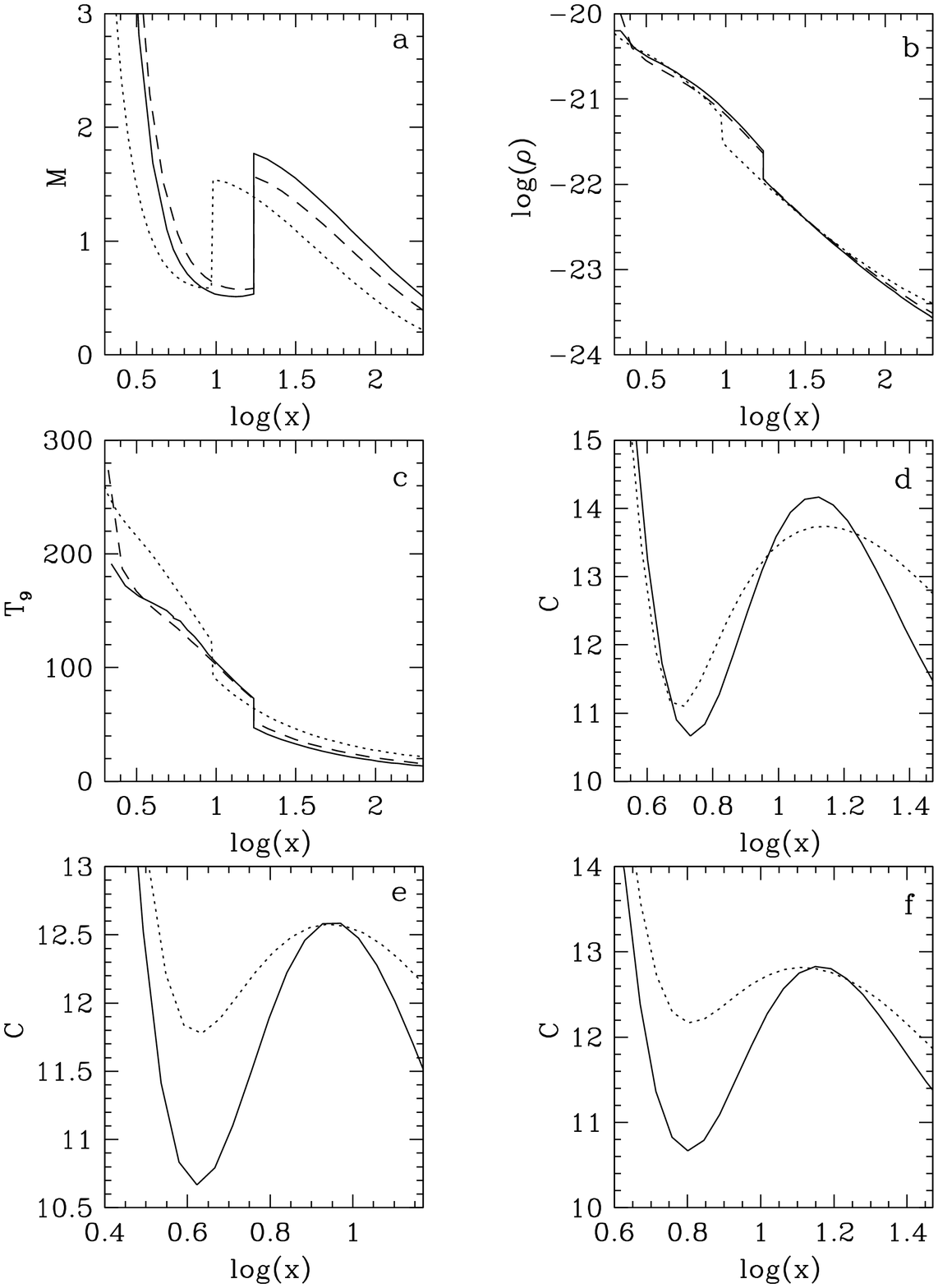}
%\plotone{f1.eps}
\label{fig2}
\end{figure}
\vskip-7.0cm
\hskip5.5cm
\noindent{\small Figure 2.
(a) Mach no., (b) density and} 

\hskip5.5cm
\noindent{\small 
(c) temperature in unit of $10^9$ for (i) $a=0.1$,} 

\hskip5.5cm
\noindent
{\small $x_o=94$, $x_i=5.4$,
$\lambda=3.3$ (solid curve);} 

\hskip5.5cm
\noindent
{\small (ii) $a=0.5$, $x_o=43$, $x_i=4.2$, $\lambda=2.8$ (dotted} 

\hskip5.5cm
\noindent
{\small curve) and
(iii) $a=-0.1$, $x_o=70$, $x_i=6.324$,}

\hskip5.5cm
\noindent
{\small $\lambda=3.3$ (dashed curve).
Shock invariant quan-}

\hskip5.5cm
\noindent
{\small tity, $C$,
for (d) parameter (i) when $x_s=17.18,$}

\hskip5.5cm
\noindent
{\small 
$9.2$, (e) parameter (ii) when $x_s=9.51,8.35$ and}

\hskip5.5cm
\noindent
{\small
(f) parameter (iii) when $x_s=17.25,13.73$.} 

\hskip5.5cm
\noindent
{\small Solid and dotted curves are for inner and outer }

\hskip5.5cm
\noindent
{\small sonic branch.
$M=10M_\odot$, $\dot{M}=1$ Eddington}

\hskip5.5cm
\noindent
{\small rate, $\gamma=4/3$.  }

\vskip1.0cm
Finally I can say, black hole angular momentum always should be considered, before making any inference 
about {\it global} disk properties. It is more important as most of black holes are thought to be rotating.
Therefore due to the rotation of a black hole not only quantitatively but also qualitatively accretion
disk properties change.

\end{document}